\documentclass[reprint,superscriptaddress, amsmath,amssymb,aps,pre,floatfix]{revtex4-1}

\usepackage[T1]{fontenc}
\usepackage[utf8]{inputenc}
\usepackage{gensymb}
\usepackage{textcomp}
\usepackage{amsmath,amsthm,mathtools}
\usepackage{amssymb}
\usepackage{graphicx}
\usepackage[english]{babel}
\usepackage[normalem]{ulem}
\usepackage{url}
\usepackage{algorithm}
\usepackage{algorithmic}
\usepackage{cancel}
\usepackage{svg}

\usepackage{color}

\usepackage{pbox} 

\usepackage{amsthm}
\newtheorem{thm}{Theorem}
\newtheorem{defn}[thm]{Definition}
\newtheorem{lemma}{Lemma}

\newtheorem{corr}{Corollary}

\usepackage{todonotes}
\usepackage{tikz}
\newcommand\copyrighttext{%
  \footnotesize This work has been submitted to the IEEE for possible publication. Copyright may be transferred without notice, after which this version may no longer be accessible.}
\newcommand\copyrightnotice{%
\begin{tikzpicture}[remember picture,overlay]
\node[anchor=south] at (current page.south) {\fbox{\parbox{\dimexpr\textwidth-\fboxsep-\fboxrule\relax}{\copyrighttext}}};
\end{tikzpicture}%
}
\begin{document}

\title{Topological theory of resilience and failure spreading in flow networks}
\author{Franz Kaiser}%
 \email{f.kaiser@fz-juelich.de}
  \affiliation{Forschungszentrum J\"ulich, Institute for Energy and Climate Research (IEK-STE), 52428 J\"ulich, Germany}
    \affiliation{Institute for Theoretical Physics, University of Cologne, K\"oln, 50937, Germany}
\author{Dirk Witthaut}%
 \email{d.witthaut@fz-juelich.de}
  \affiliation{Forschungszentrum J\"ulich, Institute for Energy and Climate Research (IEK-STE), 52428 J\"ulich, Germany}
    \affiliation{Institute for Theoretical Physics, University of Cologne, K\"oln, 50937, Germany}  
\date{\today}

\begin{abstract}
    Link failures in supply networks can have catastrophic consequences that can lead to a complete collapse of the network. Strategies to prevent failure spreading are thus heavily sought after. Here, we make use of a spanning tree formulation of link failures in linear flow networks to analyse topological structures that prevent failures spreading. In particular, we exploit a result obtained for resistor networks based on the \textit{Matrix tree theorem} to analyse failure spreading after link failures in power grids. Using a spanning tree formulation of link failures, we analyse three strategies based on the network topology that allow to reduce the impact of single link failures. All our strategies do not reduce the grid's ability to transport flow or do in fact improve it - in contrast to traditional containment strategies based on lowering network connectivity. Our results also explain why certain connectivity features completely suppress any failure spreading as reported in recent publications.
\end{abstract}

\maketitle
\copyrightnotice

\section{Introduction}\label{sec:intro}
The theory of linear flow networks provides a powerful framework, allowing to study systems ranging from water supply networks~\cite{Hwan96,diaz2016} and biological networks, such as leaf venation networks~\cite{Corson2010,Kati10,Hu2013}%
, to resistor networks~\cite{bollobas1998,kirchhoff_ueber_1847,van_mieghem_2017}, or AC power grids~\cite{Hert06,Wood14}. Failures of transportation links in these networks can have catastrophic consequences up to a complete collapse of the network. As a result, link failures in linear flow networks and their prevention are a field of active study~\cite{strake2018,kaiser_collective_2020,gavrilchenko_resilience_2019,cetinay_topological_2018,Guo17,guo_localization_2020_2,guo_localization_2020}.

The study of linear flow networks is intimately related to graph theory since most phenomena can be analysed on purely topological grounds~\cite{bollobas1998}. This connection dates back to work by Kirchhoff~\cite{kirchhoff_ueber_1847} who analysed resistor networks, and introduced several major tools that are now the basis of the theory of complex networks, such as the matrix tree theorem~\cite{kirchhoff_ueber_1847,maxwell_2010,bollobas1998}. These tools can now serve as a basis for the analysis of failure spreading in AC power grids, which can be modelled as linear flow networks based on the DC approximation~\cite{Wood14}. A substantial part of security analysis in power grids is dedicated to the study of transmission line outages since they can lead to cascading outages in a series of failures~\cite{Pour06,witthaut_nonlocal_2015,yang_small_2017}.

The topological approach to failure spreading has been exploited to demonstrate that the strength of flow rerouting after link failures decays with distance to the failing link~\cite{strake2018,kaiser_collective_2020,gavrilchenko_resilience_2019,cetinay_topological_2018}. In particular, the so-called rerouting distance based on cycles in the network has been found to predict flow rerouting very well~\cite{strake2018}. However, the analysis of flow rerouting still lacks a theoretical foundation. Here, we demonstrate that these observations made for flow rerouting may be understood based on a formalism originally developed to study current flows in resistor networks that uses spanning trees (STs) of the underlying graph. Moreover, the formalism explains recent results regarding the shielding against failure spreading in complex networks.

This publication is structured as follows; in the first section, we give an overview over the theory of linear flow networks and present an important lemma that relates the current flows in these networks to STs. In the next section, we demonstrate the analogy between such networks and AC power grids in the DC approximation and relate the ST formulation to line outages studied in power system security analysis. Finally, we show how this formulation may be used to understand why certain connectivity features inhibit failure spreading extending on recent results~\cite{Kaiser_2020}.

\begin{figure*}[t!]
    \centering
\includegraphics[width=0.8\textwidth]{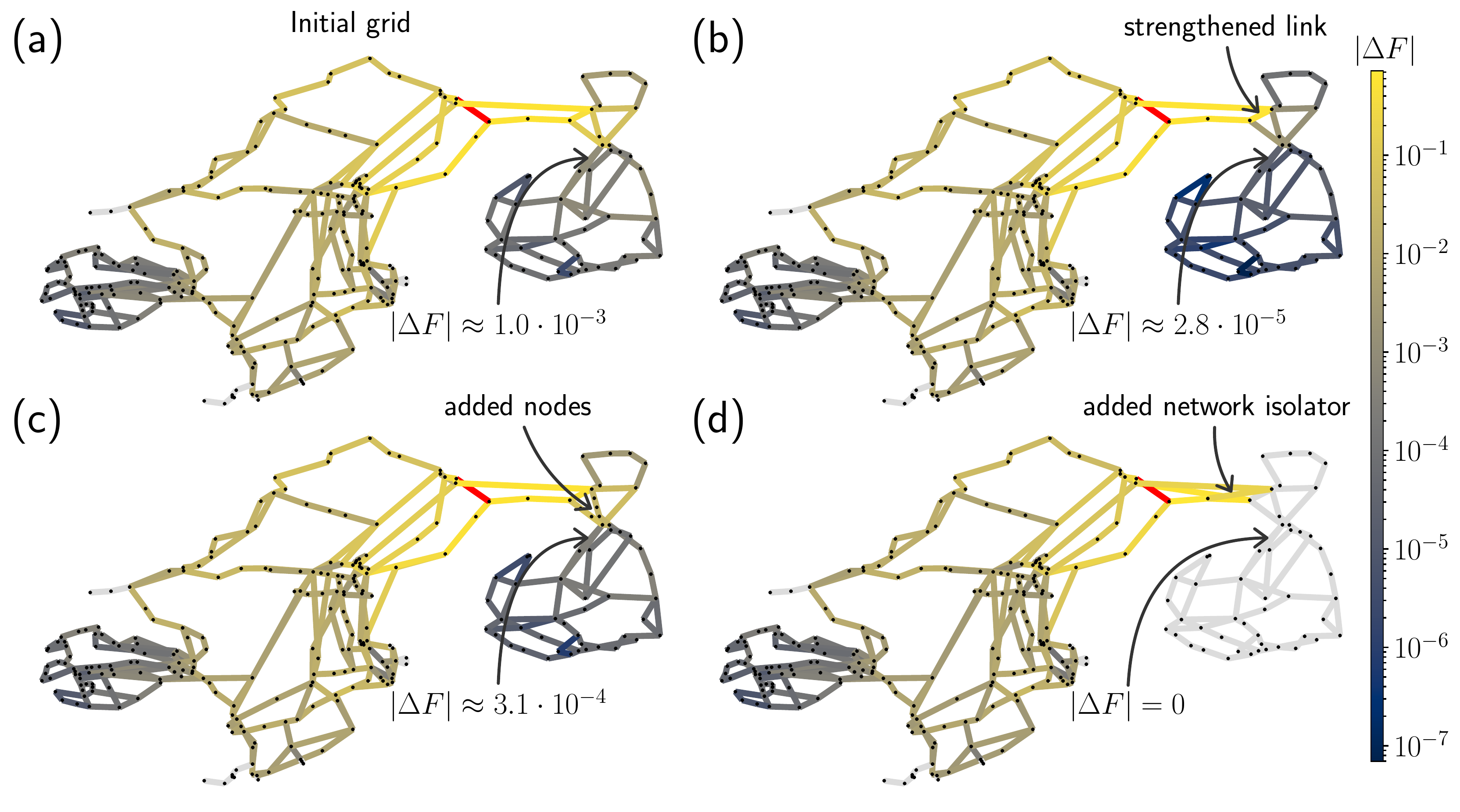}
\caption{Different methods for mitigating failure spreading in linear flow networks. (a) The failure of a single link (red) with unit flow results in flow changes $\Delta F$ (color code) throughout the Scandinavian power grid. (b) Failure spreading to Finland may be reduced by strengthening a link that horizontally separates Sweden and Finland. (c) Adding nodes, thus increasing the length of the rerouting path, reduces failure spreading to Finland as well. (d) Adding two links to construct a network isolator results in a complete vanishing of flow changes in the other part of the grid. Grid topology was extracted from the open energy system model PyPSA-Eur~\cite{horsch_2018}. } 
\label{fig:shielding_subgraphs}
\end{figure*}

\section{Fundamentals of resistor networks}
\label{sec:resistor_networks}
Resistor networks are a prime example of linear flow networks and have inspired research throughout centuries~\cite{bollobas1998,kirchhoff_ueber_1847,belevitch_summary_1962}. A resistor network can be described using a graph as follows; let $G=(E, V)$ be a connected graph with vertex set $ V=\{v_1,...v_N\}$ and $M$ edges in the edge set $E$. Then we assign a weight $w_k$ to each edge $e_k=(a,b)$ in the graph given by the inverse resistance $w_k=R_{k}^{-1}$ between its terminal vertices $a$ and $b$. If there is a potential difference $v_k=V_a-V_b$ between the terminal vertices of edge $e_k=(a,b)$, according to Ohm's law there is a current flow $i_k$ between the two vertices given by 
\begin{align}
    i_{k}=\frac{v_k}{R_k}=\frac{V_a-V_b}{R_k}.
    \label{eq:ohms_law}
\end{align}
In order to give a direction to the current flow, we assign an arbitrary orientation to each edge in the graph that is encoded by the graph's edge-node-incidence matrix $\mathbf{B}\in\mathbb{R}^{N\times M}$ defined as~\cite{bollobas1998}
\begin{align}
   B_{n,\ell} = \left\{
   \begin{array}{r l}
      1 & \; \mbox{if line $\ell$ starts at node $n$},  \\
      - 1 & \; \mbox{if line $\ell$ ends at node $n$},  \\
      0     & \; \mbox{otherwise}.
  \end{array} \right.
    \label{eq:incidence_matrix}
\end{align}
The current flows and voltages are then subject to \textit{Kirchhoff's circuit laws}~\cite{kirchhoff_ueber_1847}. The first of the laws, typically referred to as Kirchhoff's current law, at an arbitrary node $j\in  V(G)$ reads as
\begin{align*}
    \sum_{e_k\in\Lambda(j)}^M i_k=I_j.
\end{align*}
Here, $I_j\in\mathbb{R}$ is the current injected into node $j$ and $\Lambda(j)\subset E(G)$ is the set of all edges the connect to node $j$ respecting their orientation. The current law may be regarded as a continuity equation and thus states that the inflows and outflows at each node in the network have to balance with the current injections at the respective node. It may be written more compactly making use of the node-edge-incidence matrix
\begin{align}
    \mathbf{B}\mathbf{i}=\mathbf{I},
    \label{eq:current_law}
\end{align}
where $\mathbf{i}=(i_1,...,i_M)^\top\in\mathbb{R}^M$ is a vector of current flows and $\mathbf{I}=(I_1,...,I_N)^\top\in\mathbb{R}^N$ a vector of current injections. On the other hand, we can also introduce a more compact notation for Ohm's law~(\ref{eq:ohms_law}) by defining a vector of nodal voltage levels $\mathbf{V}=(V_1,...,V_N)^\top\in\mathbb{R}^N$ and a diagonal matrix of edge resistances $\mathbf{R}=\operatorname{diag}(R_1,...,R_M)\in\mathbb{R}^{M\times M}$ such that Ohm's law reads as 
\begin{align}
    \mathbf{R}\mathbf{i}=\mathbf{B}^\top \mathbf{V}.
    \label{eq:voltage_law}
\end{align}
Combining Ohm's law with Kirchhoff's current law, we arrive at the following relationship between nodal voltages $\mathbf{V}$ and nodal current injections $\mathbf{I}$
\begin{align}
    \mathbf{I}=\mathbf{B}\mathbf{R}^{-1}\mathbf{B}^\top \mathbf{V}.
    \label{eq:Poisson}
\end{align}
 This Poisson-like equation has been analysed in different contexts~\cite{strake2018,Norman97,bollobas1998}. Note that Kirchhoff's voltage law is automatically satisfied by virtue of equation~(\ref{eq:current_law}), because the resulting vector of potential differences $\mathbf{v}=\mathbf{B}^T\mathbf{V}$ vanishes along any closed cycle due to the duality between the graph's cycle space and its cut space~\cite{bollobas1998,Dies10}. In addition to that, the potential at one node may be chosen freely without affecting the result.

\begin{figure*}[t!]
    \centering
\includegraphics[width=1.0\textwidth]{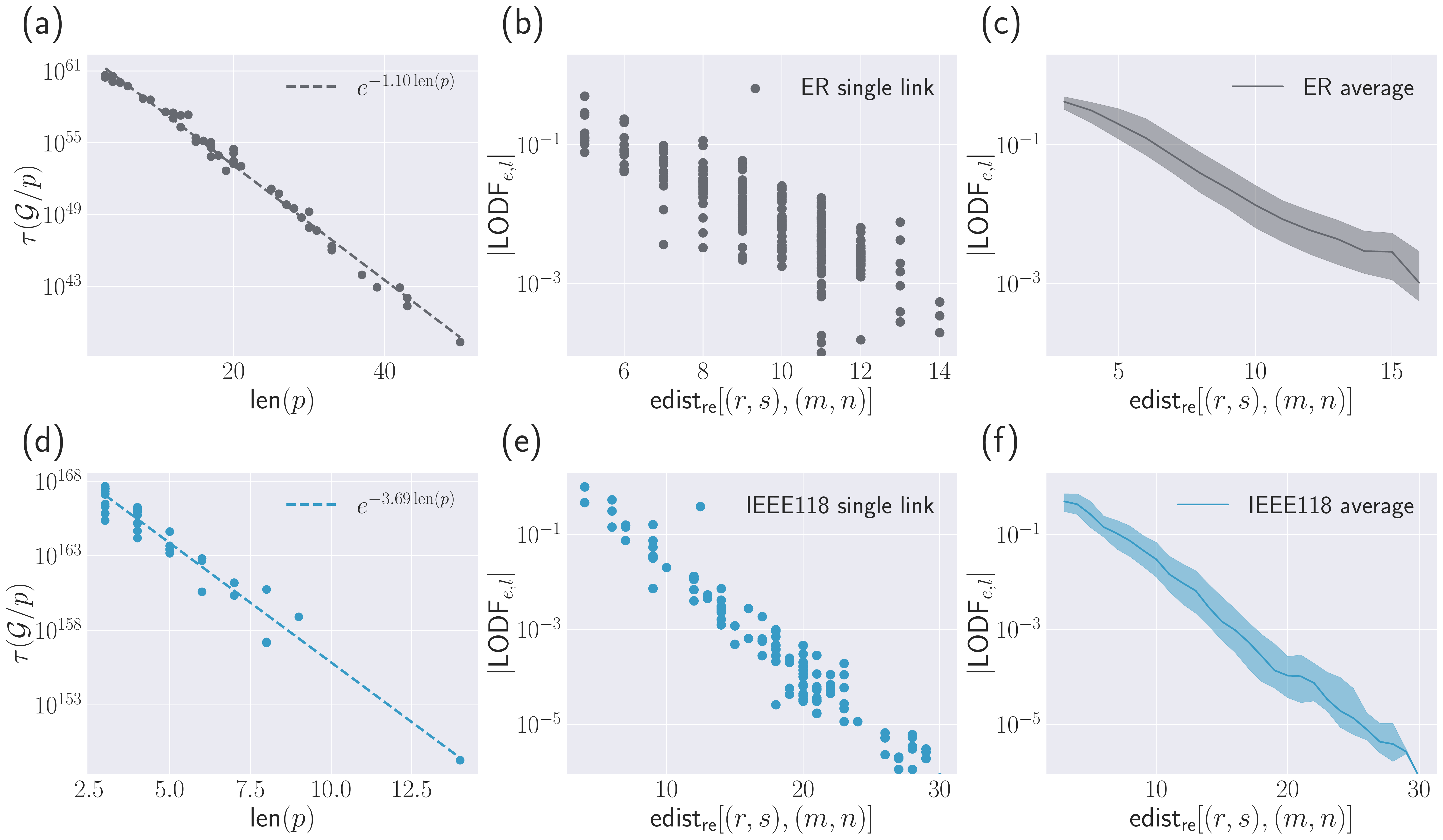}
\caption{Flow changes decay exponentially with cyclic paths in different networks. (a,d) Number of STs $\tau(G/p)$ in an Erd\H{o}s-R\'{e}nyi (ER) random graph $G(200,300)$ with $300$ edges and $200$ vertices (a) and in the power flow test case 'IEEE 118'~\cite{MATPOWER} (d) that contain a randomly chosen cyclic path $p$ (y-axis) plotted against the length of the path $\operatorname{len}(p)$ (x-axis). The number of STs decays exponentially with the length of the path, thus appearing linear on a logarithmic y-scale. (b,e) The rerouting distance scales exponentially with the LODF evaluated here for a single trigger for both grids. (c,f) The exponential scaling is preserved when averaging over all possible trigger links. Shading indicates $0.25$ and $0.75$ quantiles, line represents median. } 
\label{fig:distance_decay}
\end{figure*}

 The matrix connecting the two quantities is referred to as weighted graph Laplacian or Kirchhoff matrix $\mathbf{L}=\mathbf{B}\mathbf{R}^{-1}\mathbf{B}^\top\in\mathbb{R}^{N\times N}$ and characterises the underlying graph completely. It has the following entries~\cite{bollobas1998}
\begin{align}
    L_{mn} =  \left\{
   \begin{array}{lll}
   \displaystyle\sum \nolimits_{\ell \in \Lambda(m)} w_{\ell} &  \mbox{if } m = n; \\ [2mm]
     - w_{\ell} & \mbox{if }  m \mbox{ is connected to } n \mbox{ by } \ell.
   \end{array} \right. \label{eq:Laplacian}
\end{align}
For a connected graph, this matrix has exactly one zero eigenvalue $\lambda_1=0$ with corresponding unit eigenvector $\mathbf{v}_1=\mathbf{1}/\sqrt{N}$ such that $\mathbf{L}\mathbf{1}=0$. For this reason, the matrix is non-invertible. This is typically overcome by making use of the graph's Moore-Penrose-pseudoinverse $\mathbf{L}^\dagger$ which has properties similar to the actual inverse~\cite{Moore1920}. 

With this formalism at hand, we can in principle now determine the current on any edge given a particular injection pattern $\mathbf{I}$ and edge resistances $\mathbf{R}$. As a start, consider the situation where each edge has a unit resistance $\mathbf{R}=\operatorname{diag}(1)$  and a unit current is injected into a particular vertex $s$ and withdrawn at another one $t$ such that $\mathbf{I}=\mathbf{e}_s-\mathbf{e}_t$, where $\mathbf{e}_i=(0,...,\underbrace{1}_{i},...,0)^\top\in\{0,1\}^M$ are the unit vectors with entry one at position $i$ and zero otherwise. In this situation, the current across any edge in the graph $\ell=(a,b)$ is given by the following lemma which dates back to Kirchhoff~\cite{maxwell_2010,kirchhoff_ueber_1847} and has been popularised by Shapiro~\cite{bollobas1998,Shap87}.

\begin{lemma}
Put a one-ampere current between the vertices $s$ and $t$ of a connected, unweighted graph $G$ such that $\mathbf{I}=\mathbf{e}_s-\mathbf{e}_t$. Then the current on any other edge $(a,b)$ is given by
\begin{equation}
    i_{ab} = \frac{
     \mathcal{N}(s,a\rightarrow b,t)-\mathcal{N}(s,b\rightarrow a,t)}{\mathcal{N}},\nonumber
\end{equation}
where $\mathcal{N}(s,a\rightarrow b,t)$ is the number of STs that contain a path from $s$ to $t$ of the form $s,\ldots,a,b,\ldots,t$ and $\mathcal{N}$ is the total number of STs of the graph. \label{lem:electricallemma}
\end{lemma}
Whereas this lemma only holds for graphs where all links have unit resistances, real-world resistor networks or other types of linear flow networks are typically weighted with non-homogeneous resistances. However, the extension to weighted networks is straightforward as summarised in the following corollary.
\begin{corr}
Put a one-ampere current between the vertices $s$ and $t$ of a connected, weighted graph $G$ such that $\mathbf{I}=\mathbf{e}_s-\mathbf{e}_t$. Then the current on any other edge $(a,b)$ is given by
\begin{equation}
    i_{ab} = \frac{
     \mathcal{N}^*(s,a\rightarrow b,t)-\mathcal{N}^*(s,b\rightarrow a,t)}{\mathcal{N}^*},\label{eq:ST_LODF}
\end{equation}
where $\mathcal{N}^*=\sum_{T\in \mathcal{T}}\prod_{e\in T}w_e$ is the sum over the products of the weights $w_e$ of all edges $e\in T$ that are part of the respective spanning tree $T$ and $\mathcal{T}$ is the set of all STs in the graph.
We thus assign a weight to each ST given by the product of the weights of the edges on the ST and replace the unweighted STs in Lemma~\ref{lem:electricallemma} by weighted STs.
\end{corr}

We will demonstrate in the following sections how this lemma and corollary may be made use of to understand how failure spreading may be mitigated in linear flow networks such as AC power grids in the DC approximation.

\section{Analogy between resistor networks and power flow in electrical grids}
\label{sec:resistor_powergrids}
Importantly, the theoretical framework developed in the last section may not only be applied to resistor networks. In this section, we demonstrate how these results may be used to gain insight into the mitigation of failure spreading in power grids. 

\subsection{Modelling power grids as linear flow networks}
Most electric power transmission grids are made up of AC transmission lines and are as such governed by the non-linear AC power flow equations~\cite{Wood14}. However, the real power flow over transmission lines can be simplified to a linear flow model in what is referred to as the DC approximation of the AC power flow. This approximation is based on the following assumptions:
\begin{itemize}
    \item Nodal voltages vary little.
    \item Transmission lines are purely inductive, i.e. their resistance is negligible compared to their reactance $r_\ell\ll x_\ell,~\forall \ell \in E(G)$.
    \item Differences between nodal voltage angles $\vartheta_n,~n\in V(G)$ of neighbouring nodes $n,m$ are small $\vartheta_n-\vartheta_m\ll1$.
\end{itemize}
Typically, these assumptions are met if the power grid is not heavily loaded~\cite{Purc05}. As a result, the real power flow $F_\ell$ along a transmission line $e_\ell=(n,m)\in E(G)$ in the DC approximation depends linearly on the nodal voltage phase angles $\vartheta_n$ of neighbouring nodes
\begin{equation}
    F_\ell = b_\ell  (\vartheta_n-\vartheta_m).
\end{equation}
Here $b_\ell\approx x_\ell^{-1}$ is the line susceptance of line $\ell$. Thus, the vector of real power flow along the transmission lines in the power grid $\mathbf{F}=(F_1, ... ,F_M)^\top\in\mathbb{R}^M$ takes the role of current flow vector in the case of resistor networks. On the other hand, the nodal voltage phase angles $\mathbf{\vartheta}=(\vartheta_1,...,\vartheta_N)^\top\in\mathbb{R}^N$ take the role of the nodal voltages $\mathbf{V}$ and line weights are given by the line susceptances $b_k$ of an edge $e_k$ in correspondence with the inverse resistances $r^{-1}_k$ in the case of resistor networks. Thus, Ohm's law~(\ref{eq:voltage_law}) translates to power grids as
 \begin{align*}
     \mathbf{F}=\mathbf{B}_d\mathbf{B}^\top\mathbf{\vartheta}.
 \end{align*}
Here, $\mathbf{B}_d = \operatorname{diag}(b_1,...,b_M)\in\mathbb{R}^{M\times M}$ is the diagonal matrix of line susceptances. Again, Kirchhoff's current law~(\ref{eq:current_law}) holds and we may express it using vector quantities as follows~\cite{Wood14,strake2018}
\begin{align*}
    \mathbf{B}\mathbf{F} = \mathbf{P}.
\end{align*}
 Here, $\mathbf{P}=(P_1,...,P_N)^\top\in\mathbb{R}^N$ is the vector of nodal power injections which thus takes the role of nodal current injections $\mathbf{I}$. We summarise these equivalences in Table~\ref{tab:resistor_dc}. 

\subsection{Sensitivity factors in power grid security analysis}

  \begin{table}[tb!]
\centering
\caption{\footnotesize \label{tab:resistor_dc}Analogy between resistor networks and AC power grids in the DC approximation.}
 \begin{tabular}{lc|lc}
 \multicolumn{2}{c|}{DC approximation} &\multicolumn{2}{c}{Resistor network} \\ \hline\hline
  Power injections  &$\mathbf{P}$ & Nodal current &$\mathbf{I}$ \\
    Real power flow  & $\mathbf{F}$& Current flow&$\mathbf{i}$\\
    Nodal phase angles & $\mathbf{\vartheta}$ & Nodal Voltages  &$\mathbf{V}$\\
    Line susceptances & $b_e$ & Inverse edge resistance & $r_e^{-1}$\\
 \end{tabular}
 \end{table}
 In power grids security analysis, linear sensitivity factors are used to study and prevent line overloads which would prevent the power grids from running properly~\cite{Wood14}. One of these factors is the \textit{Power Transfer Distribution Factor} (PTDF). The PTDF$_{s,t,k}$ then quantifies the change in flow $\Delta F_k$ on line $e_k\in E(G)$ if a power $\Delta P$ is injected at node $r$ and withdrawn from node $s$. It is calculated as 
 \begin{align}
     \text{PTDF}_{r,s,k} = \frac{\Delta F_{k}}{\Delta P}.
     \label{eq:PTDF}
 \end{align}
 In addition to this factor, one typically considers the \textit{Line Outage Distribution Factor} (LODF) which measures the change in power flow on a line $e_m$ when another line $e_k$ fails~\cite{Wood14}
\begin{align}
     \text{LODF}_{m,k}=\frac{\Delta F_m}{F_k^{(0)}}.
\label{eq:lodf}
\end{align}
Here, $F_k^{(0)}$ is the flow on line $e_k$ before the outage. Mathematically, these two quantities are related as follows if $e_k=(r,s)$ is the failing link~\cite{Wood14}
\begin{align}
\text{LODF}_{m,k} = \frac{\text{PTDF}_{r,s,m}}{1-\text{PTDF}_{r,s,k}}.
    \label{eq:LODF_PTDF}
\end{align}
\begin{figure*}[t!]
    \centering
\includegraphics[width=1.0\textwidth]{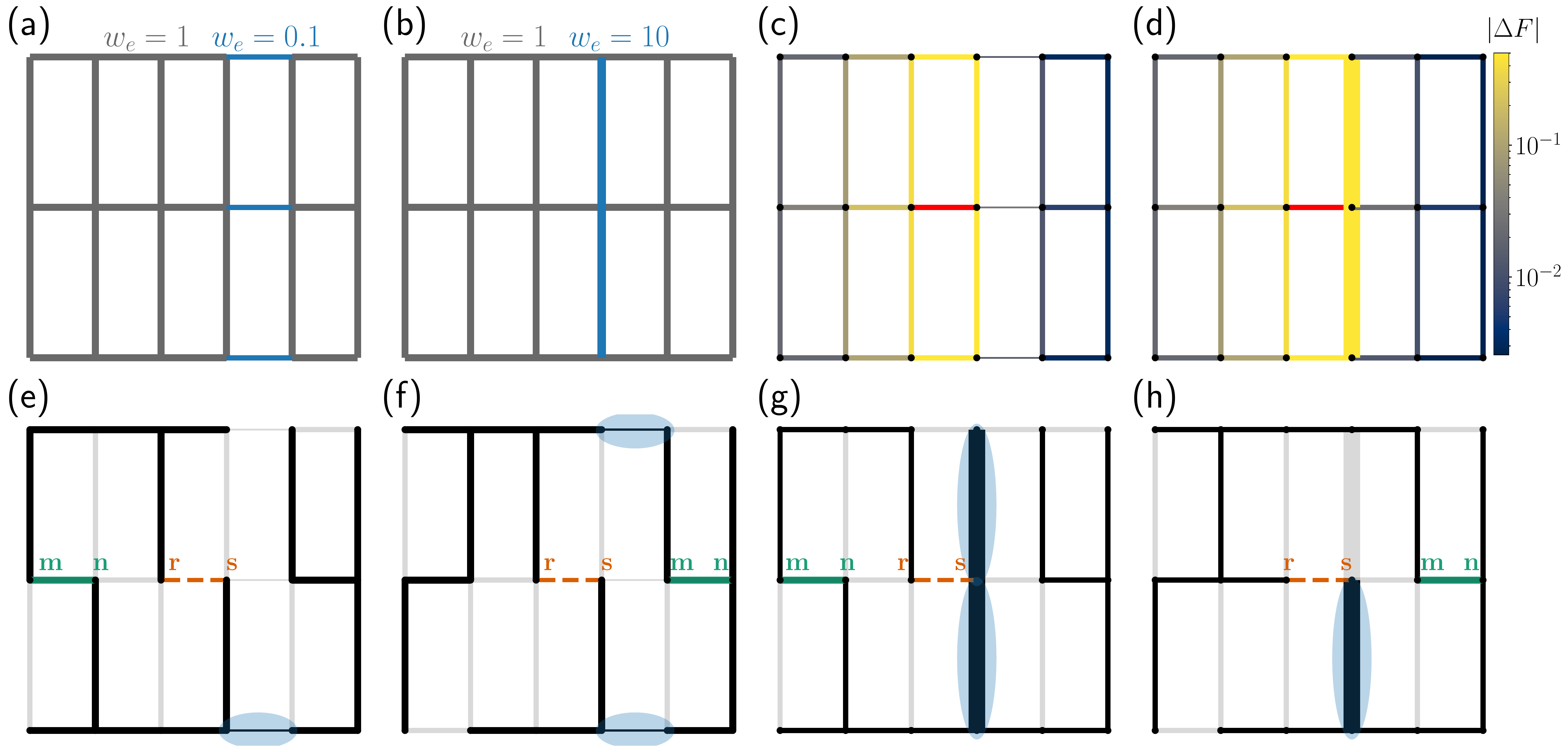}
\caption{Spanning trees (STs) may be used to explain the shielding effect of certain connectivity structures between different parts of a network. (a,b) A square grid is divided into two parts by either weakening the links connecting two parts (a, blue, $w_e=0.1$) or strengthening the links perpendicularly separating the two parts (b, blue, $w_e=10$). (c,d) For both divisions, the failure of a single link with unit flow (red) significantly reduces failure spreading to the other part of the network. 
(e-h) Different STs (black) that contain specific paths of the form
$(v_0 = r, v_1, \ldots, v_i = m, v_{i+1} = n, v_{i+2}, \ldots, v_k=s)$ used to calculate the flow changes on link $(m,n)$ for a failure of link $(r,s)$ by virtue of Eq~(\ref{eq:ST_LODF}). (e,f) For the weakly connected network shown in panel (a,c), a monitoring link in the same part (e) may lead to STs that contain only one weak link (blue shading). Thus, the contribution of this ST to the sum over all STs is much stronger than for a monitoring link in the other part, where STs have to contain at least two weak links (f, blue shading). (g,h) For the strongly connected network shown in panel (g,h), the STs contributing highest are the ones containing all edges with strong weights (g, blue shading). (h) If links $(m,n)$ and $(r,s)$ are in different parts, no ST may contain all edges with strong weights (blue shading), thus reducing failure spreading in this case.} 
\label{fig:sg_spanningtrees}
\end{figure*}

\subsection{Spanning tree description of link failures}
On the basis of the analogy between electrical grids and resistor networks developed in the last sections, we will now show how the ST formula presented in Lemma~\ref{lem:electricallemma} may be used for power systems security analysis. In the language of power grids, the lemma yields the $\text{PTDF}_{s,t,m}$ for an edge $e_m=(a,b)$ if a unit power $\Delta P$ is injected at node $r$ and withdrawn from node $s$. For this reason, the PTDF may be calculated as follows
\begin{align}
\text{PTDF}_{s,t,m}=\frac{\mathcal{N}^*(s,a\rightarrow b,t)-\mathcal{N}^*(s,b\rightarrow a,t)}{\mathcal{N}^*}.
    \label{eq:PTDF_trees}
\end{align}
Based on Eq.~\eqref{eq:LODF_PTDF} which yields the LODF expressed in terms of the PTDF, we can make use of this expression to derive an equivalent expression for the LODF. If $e_k=(r,s)$ is the failing link and $e_m=(a,b)$ the link where the flow changes are monitored, the expression based on Eq.~\eqref{eq:PTDF_trees} reads as
\begin{align}
     \text{LODF}_{m,k}&=\frac{\mathcal{N}^*(r,a\rightarrow b,s)-\mathcal{N}^*(r,b\rightarrow a,s)}{\mathcal{N}^*-(\mathcal{N}^*(r,r\rightarrow s,s)-\mathcal{N}^*(r,s\rightarrow r,s))}\nonumber\\
     &=\frac{\mathcal{N}^*(r,a\rightarrow b,s)-\mathcal{N}^*(r,b\rightarrow a,s)}{\mathcal{N}^*-\mathcal{N}^*(r,r\rightarrow s,s)}\nonumber\\
     &=\frac{\mathcal{N}^*(r,a\rightarrow b,s)-\mathcal{N}^*(r,b\rightarrow a,s)}{\mathcal{N}^*_{\backslash \{k\}}}.
    \label{eq:LODF_trees}
\end{align}
Here, $\mathcal{N}^*_{\backslash \{k\}}$ denotes the weight of all STs in the graph evaluated \emph{after} removing the edge $e_k$. We thus found an expression for the LODFs that is based purely on certain STs in the graph. This equation is the basis of our analysis of subgraphs inhibiting failure spreading which we will perform in the following sections. Note that a similar expression for the LODFs based on spanning 2-forests has recently been derived by Guo et al.~\cite{Guo17}.

\section{Mitigating failure spreading}
\label{sec:mitigating_spreading}
We have seen in the last section that the spreading of failures is studied using LODFs in power systems security analysis. To prevent large flow changes on other links after the failure of a link $e_k$ which may potentially trigger dangerous cascades of failures, it is desirable for overall power system security to keep the LODFs small. A natural question to ask is thus: Can we design or alter the network topology in such a way that LODFs stay small? Based on Equation~\eqref{eq:LODF_trees} expressing the LODF in terms of STs, this question may be addressed in a purely topological manner. In particular, we deduce three strategies to reduce the effect of failure spreading
\begin{enumerate}
    \item Fixing long paths between trigger link $e_k$ and monitoring link $e_l$ leaves only few degrees of freedom which reduces the relative contribution of the numerator in Eq.~\eqref{eq:LODF_trees}
    \item Fixing specific paths between trigger link $e_k$ and monitoring link $e_l$ can force links of large weights to be not contained in the numerator, thus reducing its relative contribution to Eq.~\eqref{eq:LODF_trees}
    \item Introducing symmetric elements between parts of the network may lead to a complete balancing between the two contributions in the numerator of Eq.~\eqref{eq:LODF_trees}
\end{enumerate}
We will address each of the strategies in the following subsections.

\subsection{The role of the rerouting distance}
\label{sec:rrdist}

With Eq.~\eqref{eq:LODF_trees} expressing LODFs using STs at hand it is intuitively clear that certain paths in the network should play an important role in predicting the overall effect of line outages. In particular, we can see immediately that for a given failing link $e_k$, the numerator in Eq.~\eqref{eq:LODF_trees} depends on the paths going through the link monitoring the flow changes $e_l$ whereas the denominator does not. Therefore, we expect the flow changes to be smaller on another link $e_m$ that has a longer minimum path going through $e_m$ and $e_k$ compared to link $e_l$. This is due to the fact that reducing the number of possible path in the sum over all STs $\mathcal{N}^*(r,a\rightarrow b,s)$ effectively reduces the number of STs by fixing a certain path. 

This intuitive idea is demonstrated to hold also quantitatively in Figure~\ref{fig:distance_decay},a,d: We illustrate that the number of STs $\tau(G/p)$ scales approximately exponentially with the length of the cyclic path contained in the STs for an unweighted Erd\H{o}s-R\'{e}nyi (ER) random graph $G(200,300)$ with $300$ edges and $200$ vertices~\cite{Erdos1960} (a) and the power flow test case 'IEEE 118'~\cite{MATPOWER,josz_ac_2016} (d). To study this scaling, we contract a cyclic path $p$ between two arbitrarily chosen edges and quantify the number of STs using Kirchhoff's matrix tree theorem~\cite{kirchhoff_ueber_1847}. The theorem states that the number of STs in a graph may be calculated using the determinant of the graph's Laplacian matrix
\begin{align*}
\tau(G)=\operatorname{det}(L_u).
\end{align*}
Here, $L_u$ is the matrix obtained from the Laplacian matrix $L$ of $G$ obtained by removing row and column corresponding to an arbitrarily chosen vertex $u\in  V(G)$. The number of STs $\tau(G/p)$ containing a path $p$ may be calculated by contracting the path in the graph and the Laplacian matrix and then taking the determinant of the resulting Laplacian. Taking the difference in the numerator of Eq.~(\ref{eq:LODF_trees}) between the path and a reversed path will in general not affect the exponential scaling since the difference of two exponentials with different exponents or different prefactors will again scale exponentially. 

\begin{figure*}[t!]
    \centering
\includegraphics[width=1.0\textwidth]{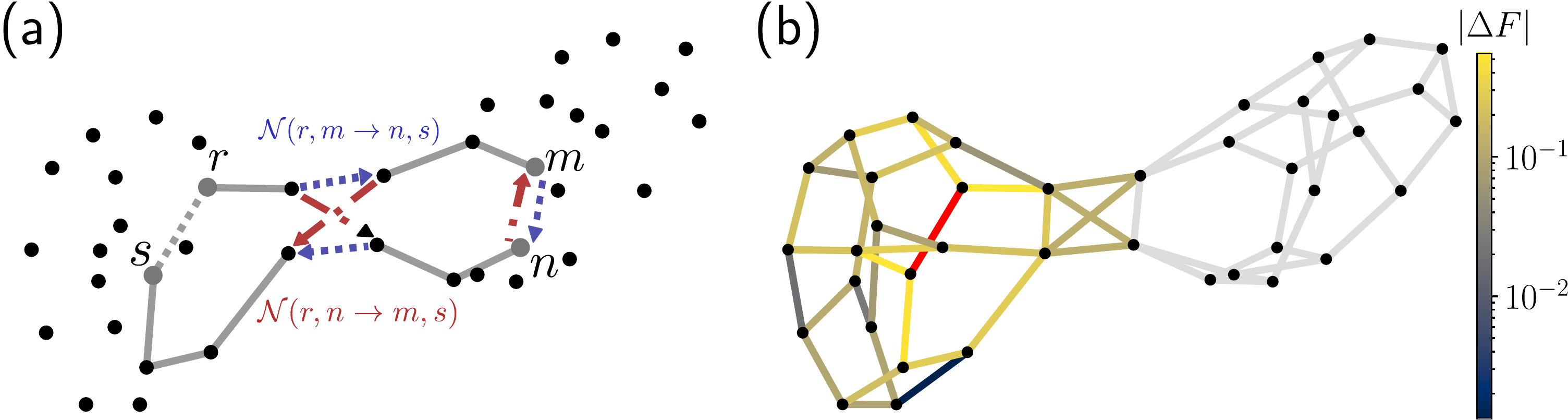}
\caption{Network isolators that lead to a complete vanishing of LODFs are created using certain symmetric paths in the network. (a) STs that contain a path starting at node $r$ and terminating at node $s$ and containing the edge $(m,n)$ (blue) or $(n,m)$ (red) have to cross the subgraph consisting of dotted, coloured edges in the centre. Since each path can contain each vertex and edge only once, each ST passing through the subgraph in one way (blue) has a counterpart passing through the subgraph in the other way (red). (b) Failure of a link (red) results in vanishing LODFs (colour code) in the part connected by a network isolator as predicted using the ST formulation of link failures.} 
\label{fig:Network_isolators}
\end{figure*}

We may thus expect an exponential decay of LODFs with the length of fixed, cyclic paths. This result complements recent progress made in the understanding of the role played by distance for failure spreading in linear flow networks. In Ref.~\cite{strake2018}, it was shown that flow changes after a link failure are not captured well by the ordinary graph distance between the failing link and the link monitoring flow changes. Instead, a different distance measure referred to as rerouting distance captures this effect much better. It is defined as follows;
\begin{defn}
A \emph{rerouting path} from vertex $r$ to vertex $s$ via the edge $(m,n)$ is a path
\begin{equation}
        (v_0 = r, v_1, \ldots, v_i = m, v_{i+1} = n, v_{i+2}, \ldots, v_k=s)\nonumber
\end{equation}
or
\begin{equation}
    (v_0 = r, v_1, \ldots, v_i = n, v_{i+1} = m, v_{i+2}, \ldots, v_k=s)\nonumber
\end{equation}
where no vertex is visited twice. The \emph{rerouting distance} between two edges $(r,s)$ and $(m,n)$ denoted by 
$${\rm edist}_{\text{re}}[(r,s),(m,n)]$$
is the length of the shortest rerouting path from $r$ to $s$ via $(m,n)$ plus the length of edge $(r,s)$. Equivalently, it is the length of the shortest cycle crossing both edges $(r,s)$ and $(m,n)$. If no such path exists, the rerouting distance is defined to be $\infty$. \label{def-reroute-dist}
\end{defn}
The rerouting distance defined this way is a proper distance metric. With the arguments made before at hand it is intuitively clear why the rerouting distance performs very well in predicting the effects of line outages. Indeed, we observe an exponential scaling of the LODFs for a given trigger link in the ER random graph Figure~\ref{fig:distance_decay},b and in the test case 'IEEE 118' (e).

\subsection{The role of strong and weak network connectivity}
\label{sec:connectivity}

Our second strategy to reduce failure spreading after link failures is based on fixing specific paths in the network in such a way that they cannot contain certain links with large weights. This way, the numerator in Equation~\eqref{eq:LODF_trees} does not contain the contribution of the links with large weights whereas the denominator does, thereby reducing the overall impact of the link failure. Note that in contrast to the last section, the fixed paths do not necessarily have to be long to prevent failure spreading. We will demonstrate this strategy for two cases: First, we use this reasoning to demonstrate that weakening the links between two parts of the network -- thus effectively dividing it into communities -- may reduce failure spreading between them. This is expected as weakly connected networks generally suppress failure spreading from one part to the other one, but this also limits the possibility of power flow between the parts. This is no longer true for the second strategy: we illustrate why also strengthening the links that separate two parts of the network horizontally reduces the impact of link failures. 

The two strategies are illustrated for a simple $3\times 6$ square grid in Figure~\ref{fig:sg_spanningtrees}. The failure of a link $e_k=(r,s)$ (dotted, orange) leads to different contribution of the numerator in Equation~\eqref{eq:LODF_trees} if the monitoring link $e_\ell=(m,n)$ (green) is contained in the same part (e) as compared to a different, weakly connected part (f) in an otherwise symmetrical situation. Note that the distance between monitoring link and trigger link is also the same in both, panels e and f. For a link in the same part, the numerator also contributes with STs containing only \textit{one} weak link (thin line, blue shading). For a trigger link located in the other part, each ST connecting trigger has to contain at least \textit{two} weak links (shaded blue). Since the contribution in the numerator is proportional to the product of all weights along the ST and the situation is otherwise symmetric, we expect a weaker LODF and thus a shielding effect if the two links are contained in different, weakly connected parts.

In panels (b) and (d), we demonstrate that strong, horizontal connections have a similar effect on failure spreading: If the monitoring link $e_\ell=(m,n)$ is contained in the same part of the network as the trigger link $e_k=(r,s)$ (g), now separated through strong connections, spanning trees connecting the two links may contain \textit{two} -- or generally: all -- strong links. For a trigger link in the other part of the network, the spanning tree connecting them can contain maximally \textit{one} -- or generally: all minus one -- strong links. Again, the term in the numerator scales with the link weights contained in the spanning trees. Therefore, we expect the effect of link failures to be stronger for links located in the same part as compared to links contained in the other part which is confirmed when simulating the failure of a single link in panel (d).

\subsection{The role of symmetry}
\label{sec:symmetry}

As a third strategy for reducing failure spreading, we suggest building networks in such a way that the terms in the numerator of Equation~\eqref{eq:LODF_trees} balance. In this case, failure spreading reduces to zero for the respective links. In order to balance the terms in the numerator of Equation~\eqref{eq:LODF_trees}, we need the spanning trees passing through the monitoring link $e_\ell=(a,b)$ in both directions to have exactly the same weight
\begin{align*}
    \mathcal{N}^*(r,m\rightarrow n,s)&=\mathcal{N}^*(r,n\rightarrow m,s)\\
    \Rightarrow\quad \sum_{T\in \mathcal{T}(r,m\rightarrow n,s)}\prod_{e\in T}w_e &=\sum_{T\in \mathcal{T}(r,n\rightarrow m,s)}\prod_{e\in T}w_e .
\end{align*}
Here, $\mathcal{T}(r,m\rightarrow n,s)$ is the set of all spanning trees containing a path of the form $(r,...,m,n,...,s)$. This equality is for example fulfilled if for each tree $T\in \mathcal{T}(r,m\rightarrow n,s)$ there is a counterpart $T\in \mathcal{T}(r,n\rightarrow m,s)$ of the same weight. This may be accomplished by introducing certain symmetric elements, referred to as \textit{network isolators}~\cite{Kaiser_2020}, into the graph as demonstrated in Figure~\ref{fig:Network_isolators}: For each ST connecting trigger link $e_k=(r,s)$ and monitoring link $e_\ell = (m,n)$ and containing a path of the form $(r,...,m,n,...,s)$ (grey and blue lines) there is an ST containing a path of the form $(r,...,n,m,...,s)$ (grey and red lines). If we compare the product of weights for a single tree $T_0\in \mathcal{T}(r,m\rightarrow n,s)$ and its counterpart $T_0^*\in \mathcal{T}(r,n\rightarrow m,s)$, such that both contain exactly the same edges except for the edges connecting the two parts, i.e., the links marked as blue and red arrows in Figure~\ref{fig:Network_isolators}, we can see that these products are equal except for the links $r_1$ and $r_2$ (red links) being contained only in $T_0$, and $b_1$ and $b_2$ (blue links) being contained only in $T_0^*$. We can thus conclude that the above equality is fulfilled, i.e., the product of weights is equal for both trees $T_0$ and $T_0^*$, if 
\begin{align*}
    b_1\cdot b_2 = r_1\cdot r_2. 
\end{align*}
In this case, a failure of link $e_k=(r,s)$ does not result in any flow changes on link $e_\ell=(m,n)$ at all. This reasoning has been generalised recently, where the concept was termed \textit{network isolators}~\cite{Kaiser_2020}. We also note that similar arguments were put forward by Guo et al~\cite{Guo17}. On general grounds, network isolators are defined as follows~\cite{Kaiser_2020}
\begin{lemma}
\label{theo:weighted}
Consider a linear flow network consisting of two parts with vertex sets $ V_1$ and $ V_2$ and assume that a single link in the induced subgraph $G( V_1)$ fails, i.e.~a link $(r,s)$ with $r,s \in  V_1$. If the adjacency matrix of the mutual connections has unit rank ${\rm rank}(\mathbf{A}_{12}) = 1$, then the flows on all links in the induced subgraph $G( V_2)$ are not affected by the failure, that is
\begin{equation}
    \begin{aligned}
   \Delta F_{m,n} \equiv 0 \quad \forall m,n \in  V_2.  
\end{aligned}\nonumber
\end{equation}
The subgraph corresponding to the mutual interactions is referred to as \textbf{network isolator}.
\end{lemma}
Note that network isolators of arbitrary size may be understood using the same reasoning as presented above for a network isolator consisting of only four links.

\begin{figure}[t!]
    \centering
\includegraphics[width=0.5\textwidth]{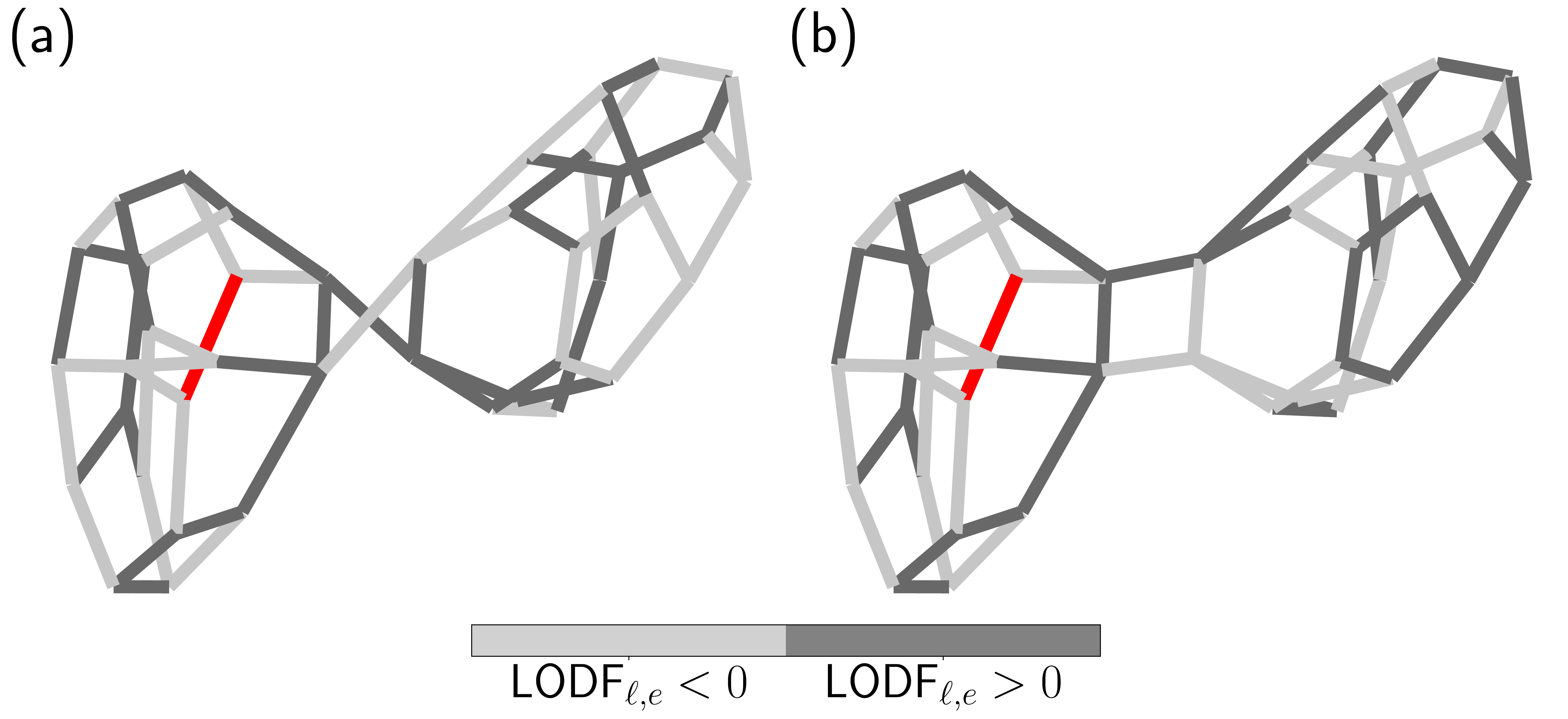}
\caption{Sign reversal of LODFs by symmetric subgraphs. (a,b) Modifying the subgraph connecting two graphs from the two parallel lines to the two crossing lines leads to a sign reversal of the LODFs in the connecting subgraphs (shades of grey). This is in line with the compensatory effect of the symmetric subgraphs used to create network isolator in Figure~\ref{fig:Network_isolators}.}
\label{fig:sign_reversal}
\end{figure}

\subsubsection{Sign reversal of flow changes}
Based on the symmetric elements -- the network isolators -- introduced in the last section, we can demonstrate yet another application of the ST formulation of link failures: We can modify the grid in such a way that the LODFs and thus the flow changes change their sign. This is again based on the symmetry of LODFs in terms of the paths $(r,...,m, n,...,s)$ and $(r,...,n, m,...,s)$. If we apply a symmetric modification such that paths of the first form are replaced by parts of the latter one, we can reverse the sign of the resulting flow changes. In particular, if we interchange the two terms appearing in the nominator of Eq.(\ref{eq:LODF_trees}= for a subset of edges, we can change the sign of the LODF for these edges
\begin{align*}
     \mathcal{N}^*(r,m\rightarrow n,s)&\rightarrow \mathcal{N}^*(r,n\rightarrow m,s)\\
     \mathcal{N}^*(r,n\rightarrow m,s)&\rightarrow \mathcal{N}^*(r,m\rightarrow n,s)\\
     \Rightarrow \text{LODF}_{\ell,k}&\rightarrow -\text{LODF}_{\ell,k}.
\end{align*}
This can achieved using a modification similar to the one shown in Fig.~\ref{fig:Network_isolators},a: If the initial network contains the subgraph indicated by dotted, blue arrows in the centre, we can revert the sign of the $\text{LODF}_{\ell,k}$ by changing this subgraph to the one indicated by red, dotted arrows. This is demonstrated in Figure~\ref{fig:sign_reversal}: Changing the subgraph in the centre connecting the two graphs from the "x"-shaped subgraph (a) to the "="-shaped subgraph (b) leads to a sign reversal of the LODFs in the second graph (shades of grey) while the magnitude of LODFs is the same in both panels. This modifications thus allows to simulatenously change the sign of all LODFs in a subgraph which may prevent overloads that are caused by flows going in a particular direction.

\section{Conclusion}
We demonstrated how a spanning tree formulation of link failures may be used to understand which topological patterns aid the mitigation of failure spreading in power grids and other types of linear flow networks. In particular, we derived and explained three strategies for reducing the effect of link failures in linear flow networks based on spanning trees. Our results offer a new understanding of previous strategies used to inhibit failure spreading in power grids and may thus help increasing power grid security.

All strategies analysed here for reducing failure spreading are based on extending -- or at least not reducing -- the network's ability to transport flows. This is in contrast to typical containment strategies in power grid security which are based on islanding the power grid, i.e. reducing the connectivity for the sake of security. We illustrated how to exploit the intimate connection to graph theory to find and analyse subgraphs that allow for improving both power grid resilience and efficiency at the same time.

Our results offer a new understanding on a graph-theoretical level of network structures that have been found to inhibit or enhance failure spreading. We illustrated the fruitful approach of analysing failure spreading in power grids by using spanning trees for several subgraphs but are confident that other subgraphs for enhancing or inhibiting failure spreading may be unveiled using this formalism.

\section*{Acknowledgments}
 We gratefully acknowledge support from the German Federal Ministry of Education and Research (grant no. 03EK3055B) and the Helmholtz Association (via the joint initiative ``Energy System 2050 -- A Contribution of the Research Field Energy'' and the grant no.~VH-NG-1025). 

\bibliography{bibliography}

\end{document}